# A Noninvasive and Dispersive Framework for Estimating Nonuniform Conductivity of Brain Tumor in Patient-Specific Head Models

YOSHIKI KUBOTA[1], YOSUKE NAGATA[1], MANABU TAMURA[2], AKIMASA HIRATA[1]

[1]Department of Electrical and Mechanical Engineering, Nagoya Institute of Technology, Nagoya 466-8555, Japan
[2]Faculty of Advanced Techno-Surgery, Institute of Advanced Biomedical Engineering and Science, Graduate School of Medicine, Tokyo Women's Medical University, Tokyo 162-0054, Japan

CORRESPONDING AUTHOR: Y. KUBOTA (e-mail: kubota.yoshiki@nitech.ac.jp).

This work was supported by the Ministry of Internal Affairs and Communications, Japan (Grant Number JPMI10001)

**ABSTRACT** We propose a noninvasive and dispersive framework for estimating the spatially nonuniform conductivity of brain tumors using MR images. The method consists of two components: (i) voxel-wise assignment of tumor conductivity based on reference values fitted to the Cole–Cole model using empirical data from the literature and (ii) fine-tuning of a deep learning model pretrained on healthy participants A total of 67 cases, comprising both healthy participants and tumor patients and including 9,806 paired T1- and T2-weighted MR images, were used for training and evaluation. The proposed method successfully estimated patient-specific conductivity maps, exhibiting smooth spatial variations that reflected tissue characteristics, such as edema, necrosis, and rim-associated intensity gradients observed in T1- and T2-weighted MR images. At 10 kHz, case-wise mean conductivity values varied across patients, ranging from 0.132 to 0.512 S/m in the rim (defined as the region within 2 mm of the tumor boundary), from 0.132 to 0.608 S/m in the core (the area inside the rim), and from 0.141 to 0.542 S/m in the entire tumor. Electromagnetic simulations for transcranial magnetic stimulation in individualized head models showed substantial differences in intratumoral field distributions between uniform assignments and the proposed nonuniform maps. Furthermore, this framework demonstrated voxel-wise dispersive mapping at 10 kHz, 1 MHz, and 100 MHz. This framework supports accurate whole-brain conductivity estimation by incorporating both individual anatomical structures and tumor-specific characteristics. Collectively, these results advance patient-specific EM modeling for tumor-bearing brains and lay the groundwork for subsequent microwave-band validation.

## I. INTRODUCTION

Precision medicine has emerged as a central paradigm in contemporary healthcare, aiming to optimize diagnosis and treatment based on individual-specific characteristics [1], [2], [3]. In the domain of neuromodulation, particularly brain stimulation techniques such as transcranial magnetic stimulation (TMS) and temporal interference stimulation (TIS), which typically operate in the kHz frequency band, computational models replicating human anatomical structures play a vital role in estimating the electric field induced in target brain regions [4], [5], [6], [7]. These models treat biological tissues as volumetric conductors, enabling the simulation of field distributions within the brain. However, as the accuracy of electric field simulations heavily depends on the precision of conductivity assignment to biological tissues, accurately estimating tissue conductivity is essential for reliable modeling [8], [9], [10].

In clinical settings, precise localization of functional regions, such as the motor and language cortices, is critical when planning the surgical resection of brain tumors, particularly when these regions lie near the lesion [11], [12]. For this purpose, noninvasive brain stimulation techniques, including TMS and TIS, have been integrated with electric field simulations to support preoperative functional brain mapping [13], [14], [15]. The clinical utility of this approach has been demonstrated, and reliably predicting stimulation-induced brain responses requires a head conductivity model that captures patient-specific properties, including those of tumors.

Despite the central role of conductivity modeling, conventional segmented head models [16], [17], [18] present two primary limitations: (1) they do not account for intra-tissue spatial heterogeneity, assigning a single, isotropic conductivity to each tissue type, and (2) the electrical conductivity of brain tumors remains largely uncharacterized due to a lack of empirical measurements and noninvasive estimation tools. This is particularly problematic because tumor tissues often exhibit complex internal structures, such as necrosis, edema, and heterogeneous cell densities, which may influence the electric field distribution in a spatially dependent manner.

To address the first limitation, we previously developed a deep learning framework, CondNet-TART, for estimating personalized head conductivity distribution from T1- and T2-weighted images [19]. This model was pretrained on a cohort of healthy participants from the IXI dataset [20], demonstrating high accuracy in estimating spatially non-uniform conductivities in normal tissues [19]. However, a key challenge is that the model has not yet been applied to cases involving brain tumors, where conductivity remains poorly characterized owing to the scarcity of reference data and expert annotations.

The second major limitation relates to the constraints of existing conductivity estimation techniques. Electrical property tomography (EPT) enables noninvasive mapping of whole-brain conductivity [21], [22], [23], [24]; however, it requires specialized MRI pulse sequences and is highly sensitive to imaging noise. Moreover, EPT-derived tissue conductivities are only valid within the MHz frequency range of the imaging sequence, limiting their applicability to simulations at other frequencies. Conversely, direct conductivity measurements require invasive intraoperative procedures [25], [26], [27], [28], and the heterogeneous, variable nature of tumor tissues makes it challenging to acquire comprehensive conductivity maps across the entire tumor. Therefore, there is a critical need for a noninvasive, dispersive method to estimate brain tumor conductivity.

In this study, we propose a noninvasive and dispersive framework for estimating the electrical conductivity of brain tumors using T1- and T2-weighted MR images. The proposed approach was then validated at 10 kHz, 1 MHz, and 100 MHz, thereby covering the MHz frequency range relevant for RF thermal therapy. To our knowledge, this is the first study to estimate nonuniform tumor conductivity distributions directly from standard anatomical MR images without the need for specialized imaging sequences or manual annotations.

## II. MATERIALS AND METHODS

### A. OVERVIEW OF THE PROPOSED METHOD

Fig. 1 illustrates a flowchart of the proposed method. The approach integrates two complementary tumor components: (i) pixel-wise assignment of tumor conductivity based on frequency-adjusted reference data from the literature and (ii) fine-tuning of a pretrained model to improve the accuracy of whole-brain conductivity estimation. These components were implemented through a two-stage training process in CondNet-TART: pretraining on healthy participants, followed by fine-tuning on tumor cases. The second component addresses this limitation by incorporating physics-informed supervision derived from the adjusted reference values. Simultaneously, it mitigates the limitation of uniform tissue conductivity by enabling the model to learn spatially heterogeneous conductivity patterns within tumors, an ability that is not present in the original model trained exclusively on healthy participants. The use of a frequency-adjusted tumor conductivity map, informed by the Cole–Cole model [29], ensures biophysical consistency and supports model generalization across a range of frequencies, overcoming a key limitation of EPT. Ultimately, this integration facilitates the generation of patient-specific conductivity maps that accurately capture both individual anatomical structures and intratumoral heterogeneity.

### B. DATASET DESCRIPTION

The proposed framework for brain tumor conductivity estimation was trained using a three-step process across two stages: (1) segmentation of healthy participants, (2) conductivity estimation for healthy participants, and (3) conductivity estimation for tumor cases. Healthy participant data were sourced from the IXI dataset [20], whereas tumor cases were collected from clinical datasets at Tokyo Women's Medical University Hospital. A total of 67 cases were used in this study, comprising 9,806 paired T1- and T2-weighted MR image sets. The number of data samples used in each training step is summarized in

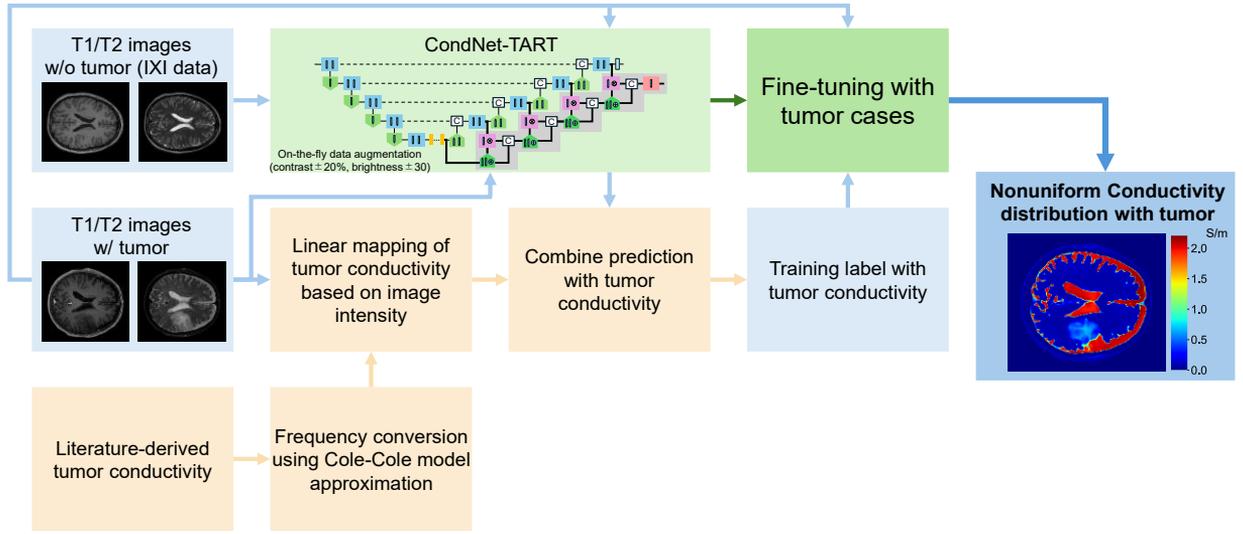

**FIGURE 1.** Flowchart of the proposed method.

TABLE 1, and the characteristics of the tumor cases are detailed in TABLE 2. For the tumor cases, cross-validation was performed using 10 of the 12 cases for training and the remaining two cases for evaluation.

This study was approved by the Institutional Review Boards of Tokyo Women's Medical University (approval no. 2022-0112) and Nagoya Institute of Technology (approval no. 2024-26). This study was conducted in accordance with the principles of the Declaration of Helsinki. Written informed consent was waived owing to the retrospective nature of the study, and information regarding the study was disclosed on the institutional website to allow patients to opt out.

**TABLE 1.** Overview of datasets for each task. Values indicate the number of participants or cases used for each task, with the number of paired T1- and T2-weighted image sets shown in parentheses.

| Step | Train | Test |
|---|---|---|
| Segmentation | 40 (5889) | |
| Conductivity Estimation for healthy participants | 10 (1471) | 10 (1494) |
| Conductivity Estimation for tumor cases | | 12 (1752) |
| Total | 67 (9806) | |

**TABLE 2.** Characteristics for tumor cases.

| Case No. | Age | Sex | Tumor Location | Tumor Type | WHO Grade |
|---|---|---|---|---|---|
| 1 | 39 | M | R F, P | Anaplastic oligo-astrocytoma | 3 |
| 2 | 65 | M | L F, P | Anaplastic oligodendroglioma | 3 |
| 3 | 40 | F | L F, SMA | Oligo-astrocytoma | 2 |
| 4 | 29 | M | L F, SMA | Anaplastic oligo-astrocytoma | 3 |
| 5 | 33 | M | R F, SMA | Oligo-astrocytoma | 2 |
| 6 | 34 | F | L P | Oligodendroglioma | 2 |
| 7 | 53 | M | R F | Glioblastoma | 4 |
| 8 | 45 | M | L F | Anaplastic astrocytoma | 3 |
| 9 | 53 | M | R F, T | Anaplastic oligodendroglioma | 3 |
| 10 | 51 | M | R F | Anaplastic astrocytoma | 3 |
| 11 | 44 | F | L SMA | Anaplastic astrocytoma, IDH-mutant | 3 |
| 12 | 43 | M | R F | Anaplastic astrocytoma, IDH mutated | 3 |
| Median (Range) | 43.5 (29-65) | M | - | - | 3 (2-4) |

M, male; F, female; L, left; R, right; F, frontal lobe; P, parietal lobe; T, temporal lobe; SMA, supplementary motor area

## C. PREPROCESSING AND PRETRAINING

As part of preprocessing, median filtering was applied to both T1- and T2-weighted MR images to reduce image noise. To further minimize the influence of outliers, each image was normalized by scaling the pixel intensities according to the 99.9th percentile value.

Segmentation labels were generated using a semi-automatic method based on FreeSurfer [30]. For each label, an isotropic and uniform tissue conductivity value was assigned using the fourth-order Cole–Cole model [29], assuming a frequency of 10 kHz relevant to TMS applications. The specific conductivity values used are listed in TABLE 3. Each voxel's conductivity was normalized to the range $[0, 1 - \tau]$, where $\tau$ was set to 0.1.

On-the-fly data augmentation was applied during pretraining for both the segmentation and conductivity estimation tasks. Specifically, we introduced random contrast variations of ±20% and brightness shifts of ±30 units [31], [32]. All images were normalized to the range $[0, 1]$ before being fed into the model.

The CondNet-TART model used in both the pretraining and fine-tuning stages is based on the U-Net architecture [33] and incorporates transformer modules [34], attention gates [35], and residual connections [36] to enhance its representational capability. Additionally, by leveraging transfer learning [37], [38], the model achieves an accurate conductivity estimation even when trained on a limited number of cases [19].

**TABLE 3. Human tissue conductivity values [S/m] based on the Cole–Cole model at 10 kHz, 1 MHz, and 100 MHz.**

| Tissue | Conductivity | | |
|---|---|---|---|
| | 10 kHz | 1 MHz | 100 MHz |
| Blood | 0.700 | 0.822 | 1.230 |
| Bone (canc.) | 0.080 | 0.090 | 0.170 |
| Bone (cort.) | 0.020 | 0.024 | 0.064 |
| Cerebellum | 0.130 | 0.180 | 0.790 |
| CSF | 2.000 | 2.000 | 2.000 |
| Dura | 0.500 | 0.503 | 0.740 |
| Fat | 0.040 | 0.044 | 0.068 |
| GM | 0.100 | 0.142 | 0.560 |
| Mucous tissue | 0.070 | 0.103 | 0.145 |
| Muscle | 0.340 | 0.503 | 0.708 |
| Skin | 0.100 | 0.140 | 0.490 |
| V. Humor | 1.500 | 1.500 | 1.500 |
| WM | 0.070 | 0.102 | 0.324 |

## D. CONSTRUCTION OF CONDUCTIVITY MAPS AND FINE-TUNING

To construct a tumor-specific conductivity map, we used literature-derived values for model training. A tumor mask was first generated from T1- and T2-weighted MR images. Conductivity values measured at 3 kHz–1 MHz were obtained from a previous study [28] and fitted using the Cole–Cole model [29]. To extend the frequency range beyond 1 MHz, an additional data point at 128 MHz was incorporated. Its value (1.2 S/m), reported as the maximum conductivity by EPT [22], was converted into an equivalent magnitude based on the relationship between the values reported in [26] and [28], and then appended to the literature-derived dataset. The combined set of values (3 kHz–128 MHz) was subsequently fitted using a second-order Cole–Cole model [29], as shown in Equation (1).

$$\sigma(f) = \sigma_0 + \sum_{i=1}^{2} \frac{\Delta\sigma_i}{1 + (j2\pi f \tau_i)^{1-\alpha_i}} \tag{1}$$

where $\sigma_0$ is the static conductivity, $\Delta\sigma_i$ is the dispersion magnitude, $\tau_i$ is the relaxation time, and $\alpha_i$ is the broadening parameter. The fitted curve and resulting parameter values are shown in Fig. 2 and TABLE 4, respectively.

Next, the conductivity values from multiple prior studies [25], [26], [27] were converted to conductivity using the frequency ratio from Equation (1) to ensure consistency with our electric field simulations. A linear mapping was then applied to assign conductivity values within the tumor mask, ranging from 0.1–0.6 S/m at 10 kHz, 0.25–1.05 S/m at 1 MHz, and 0.3-1.2 S/m at 100 MHz, based on the pixel intensities of the T1- and T2-weighted MR images, as shown in Equation (2).

$$\sigma_{tumor} = C_{min} + (C_{max} - C_{min}) \times \{(1 - \alpha)(1 - \tilde{I}_{T1}) + \alpha\tilde{I}_{T2}\} \tag{2}$$

Here, $\tilde{I}_{T1}$ and $\tilde{I}_{T2}$ represent the pixel intensities of the T1- and T2-weighted images, respectively, normalized across all cases using the 5th to 95th percentile range. Additionally, $C_{min}$ and $C_{max}$ represent the lower and upper bounds of the tumor conductivity range at the target frequency, respectively. The parameter $\alpha$ is a weighting factor between the normalized T1 and T2 image intensities, and in this study, $\alpha$ was set to 0.7. These conductivity values were normalized using the same scaling parameters applied to the normal tissues to ensure consistency across the entire volume.

The resulting tumor conductivity values were combined with the output of the pretrained CondNet-TART model for nontumor regions to provide supervision during tumor case training. The same loss function used in the pretraining stage, defined in Equation (3), was employed in this training phase.

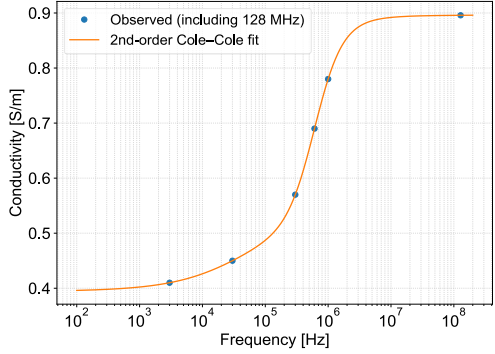

**FIGURE 2.** Literature-derived conductivity values and the fitted curve obtained using the second-order Cole–Cole model.

$$Loss = \frac{1}{n}\sum_{i \in N} l_i + \lambda \cdot \frac{1}{m}\sum_{i \in T} l_i,$$

$$l_i = 0.95|\hat{y}_i - y_i| + 0.05\log(\varepsilon + |\hat{y}_i - y_i|)$$

$$(3),$$

where $\hat{y}_i$ and $y_i$ are the estimated and supervisory conductivity values at pixel $i$, respectively. $\varepsilon$ is a small positive constant (set to $1 \times 10^{-6}$) used to ensure numerical stability in the logarithmic term. $N$ and $T$ denote the sets of pixels in non-tumor and tumor regions, respectively, with $n = |N|$ and $m = |T|$. The parameter $\lambda$ is a weighting factor used to balance the tumor region's contribution during training, was set to 3 in this study.

**TABLE 4.** Fitted parameters of the second-order Cole–Cole model for tumor conductivity.

| Parameter | Value |
|---|---|
| $\sigma_0$ | 0.3941 |
| $\Delta\sigma_1$ | 0.1241 |
| $\tau_1$ | $3.87 \times 10^{-6}$ |
| $\alpha_1$ | 0.3984 |
| $\Delta\sigma_2$ | 0.3783 |
| $\tau_2$ | $2.59 \times 10^{-7}$ |
| $\alpha_2$ | 0.0088 |

### E. ELECTROMAGNETIC SIMULATION

The induced electric field **E** was computed based on the estimated spatial distribution of conductivity using the scalar potential finite difference method[39]. The scalar potential $\psi$ satisfies Equation (4):

$$\nabla \cdot (\sigma\nabla\psi) = -\nabla \cdot \left(\sigma\frac{dA}{dt}\right)$$

$$(4),$$

where $\sigma$ denotes the conductivity distribution, and $A$ represents the magnetic vector potential of the applied magnetic field. The electric field **E** was then derived from $\psi$ and $A$ as defined in Equation (5):

$$\mathbf{E} = -\nabla\psi - \frac{dA}{dt}$$

$$(5).$$

This quasi-static formulation is appropriate for frequencies up to 10-100 MHz range, where capacitive displacement currents are negligible and tissue permittivity can be ignored [39], [40]. Accordingly, only the conduction currents were considered, and the electric field depended solely on the conductivity distribution. Equation (4) was solved numerically using the scalar potential finite-difference method [40], [41] with computational acceleration provided by the successive over-relaxation method [42].

### F. EXPOSURE SCENARIO

To explore the potential utility of the proposed method for personalized electric field modeling in neurostimulation research, we applied the estimated conductivity maps to the TMS simulations. These simulations targeted the motor cortex on the tumor-affected side of the brain in the evaluation cases, with the coil positioned at either C3 or C4, in accordance with the 10–10 international system [43]. The TMS coil was modeled using a thin-wire approximation, with outer and inner diameters of 9.7 cm and 4.7 cm, respectively. A dominant frequency of the TMS coil was set to 10 kHz.

### G. EXPERIMENTAL SETTINGS

The training process was divided into two stages: pretraining on healthy participants and fine-tuning on the tumor cases. During pretraining, the segmentation and conductivity estimation tasks were trained independently. Fine-tuning was then performed on tumor cases using adjusted learning rates and batch sizes. Detailed training parameters for each stage are presented in TABLE 5. In the CondNet-TART architecture (Fig. 1), the conductivity estimation layers are indicated by the gray-shaded region, whereas the remaining components correspond to the segmentation layers.

To evaluate the performance of the proposed method, we compared it with two baseline settings: original, where the model was trained only on healthy participants without fine-tuning, and uniform, where the model was fine-tuned on tumor cases using a fixed conductivity value of 0.35 S/m as supervision. In addition, to demonstrate the dispersion capability of the proposed method, the conductivity distributions estimated at 10 kHz, 1 MHz, and 100 MHz were compared using the same evaluation metrics.

The conductivity estimation performance was assessed using line plots and histograms of the estimated distributions. The tumor region was subdivided into the rim, core, and whole subregions. The rim was defined as the 2 mm inner margin from the tumor boundary, the core as the area inside the rim, and the whole as the entire tumor region. For each subregion, the mean conductivity was calculated per case. Then, across these case-wise means, we computed the standard deviation and range to assess the inter-case variability. In addition, the spatial gradient of conductivity was calculated within each case, and its mean, standard deviation, and range were analyzed across all cases. To exclude the influence of cerebrospinal fluid (CSF), only voxels with conductivity values below 1.5 S/m were included in these computations. For the electric field simulations, representative coronal slices were visually examined to compare the tumor region field distributions across different model settings.

The conductivity estimation was performed on a workstation with an Intel Xeon w5-2455 CPU at 3.07 GHz, 512 GB of RAM, dual NVIDIA RTX A5500 GPUs, and Windows Server 2022 Standard. The implementation was performed using Python 3.8, PyTorch 2.3.1 and CUDA Toolkit 11.8. Electric field simulations were performed on a separate workstation equipped with an Intel Xeon Gold 6250 CPU at 3.90 GHz, 192 GB of RAM, and Ubuntu 20.04.1 LTS using MATLAB R2022b and Fortran 90.

**TABLE 5.** Training parameters for each step.

| Stage | Step | Learning Rate * | Batch Size | Epochs | Loss Function |
|---|---|---|---|---|---|
| Pre-training | Segmentation | 0.001/ - | 32 | 50 | 0.5×Tversky + 0.5×BCE |
| | Conductivity Estimation for healthy participants | 0.0001/ 0.001 | 32 | 50 | MAE + log error (Equation 3) |
| Fine-tuning | Conductivity Estimation for tumor cases | 0.0005/ 0.0001 | 16 | 50 | MAE + log error + class balance (Equation 3) |

*Segmentation Layer/ Estimation Layer

## III. RESULTS

### A. CONDUCTIVITY ESTIMATION
Fig. 3 illustrates examples of conductivity maps from the original, uniform, and proposed methods. Fig. 4 presents the corresponding line and box plots derived from the same coronal slices shown in Fig. 3. TABLE 6 summarizes the estimated conductivity values within the tumor region across all 12 cases. The proposed method produced visibly heterogeneous conductivity patterns in the tumor region. Notably, it demonstrated the lowest spatial gradients in both the rim and core regions compared to the other methods, indicating a smoother and more consistent conductivity distribution.

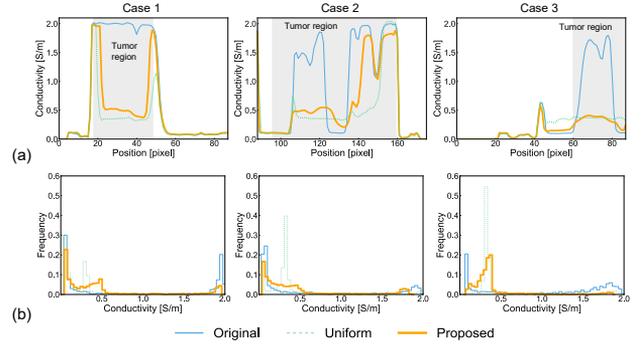

**FIGURE 4.** Detailed analysis of conductivity distributions at 10 kHz within tumor regions. (a) Line profile of tissue conductivity along the gray dashed line indicated on the conductivity map in Fig. 2 (b) Histogram of conductivity values within the tumor region, as defined by the white mask shown in Fig. 2.

### B. ELECTROMAGNETIC SIMULATION
Fig. 5 presents examples of the electric field simulation results for the three cases shown in Fig. 3, with TMS targeting the motor cortex on the tumor-affected side. While the overall field distribution outside the tumor remains largely consistent across methods, differences are evident within the tumor region. In case 2, the original method produced several localized high field intensities, whereas in case 3, it yielded lower field intensity at the tumor center compared to the other methods.

### C. DISPERSIVE CHARACTERISTICS
Fig. 6 illustrates examples of conductivity maps generated at 10 kHz, 1 MHz, and 100 MHz using the proposed method. TABLE 7 summarizes the mean conductivity values within the tumor region across all 12 cases. The conductivity estimated at 1 MHz and 100 MHz demonstrated heterogeneous distributions among the tumor subregions, similar to those observed at 10 kHz, whereas the mean values were elevated relative to the 10 kHz estimates.

## IV. DISCUSSION
In this study, we proposed a novel method for noninvasively estimating the electrical conductivity distribution of brain tumors using MRI. The framework is dispersive and applicable across the kHz-to-MHz frequency range. A key feature of the proposed method is its ability to reproduce spatial heterogeneity in the conductivity of unseen tumor cases, relying solely on T1- and T2-weighted MR images.

Figs. 3 and 4 show that the original method frequently overestimates conductivity in tumor regions, with values resembling those of CSF. This is likely due to the misclassification of tumor pixels that exhibit T1 and T2 intensities similar to those of the CSF. In contrast, the uniform method suppresses such high conductivity values and maintains a globally low and homogeneous distribution. However, similar to its training labels, it fails to capture pathological features such as edema and necrosis owing to its uniformity. The proposed method demonstrated smooth conductivity distributions that visually reflected the structural gradients within the tumors and partially captured the pathological changes. As shown in TABLE 6, it achieved lower conductivity gradients in the rim, core, and whole-tumor regions, indicating smoother spatial distributions. This suggests that the proposed approach effectively balances the preservation of anatomical variations with smoother spatial transitions. In particular, the reduced conductivity gradients observed with the proposed method indicate more physiologically plausible transitions across tissue boundaries, which may enhance the realism of the electric field simulations. In addition, because it leverages a pretrained model, the proposed method generates consistent results in non-tumor regions, such as the CSF and skull, comparable to the original method. These results underscore the potential of our unified framework for estimating whole-head conductivity, regardless of the presence of a tumor.

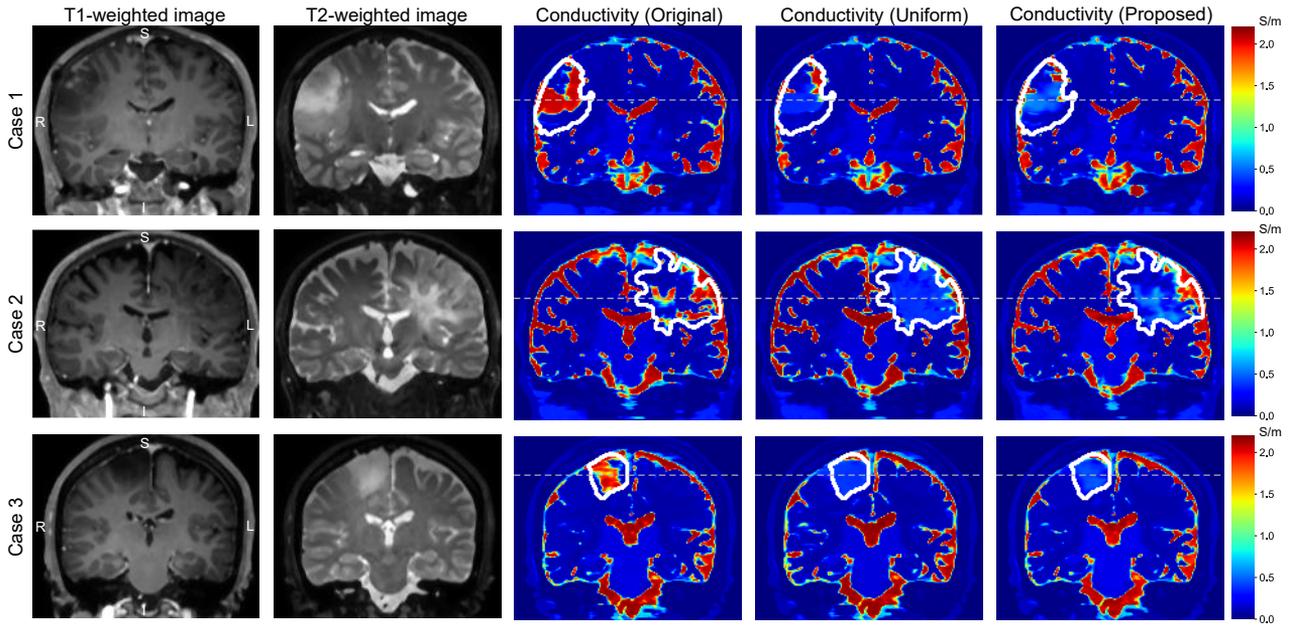

**FIGURE 3.** Estimated electrical conductivity distribution at 10 kHz in brain tumor cases. The left two columns display coronal views of the T1-weighted and T2-weighted images, while the right three columns show the estimated conductivity maps generated by the original, uniform, and proposed methods. White masks denote tumor regions.

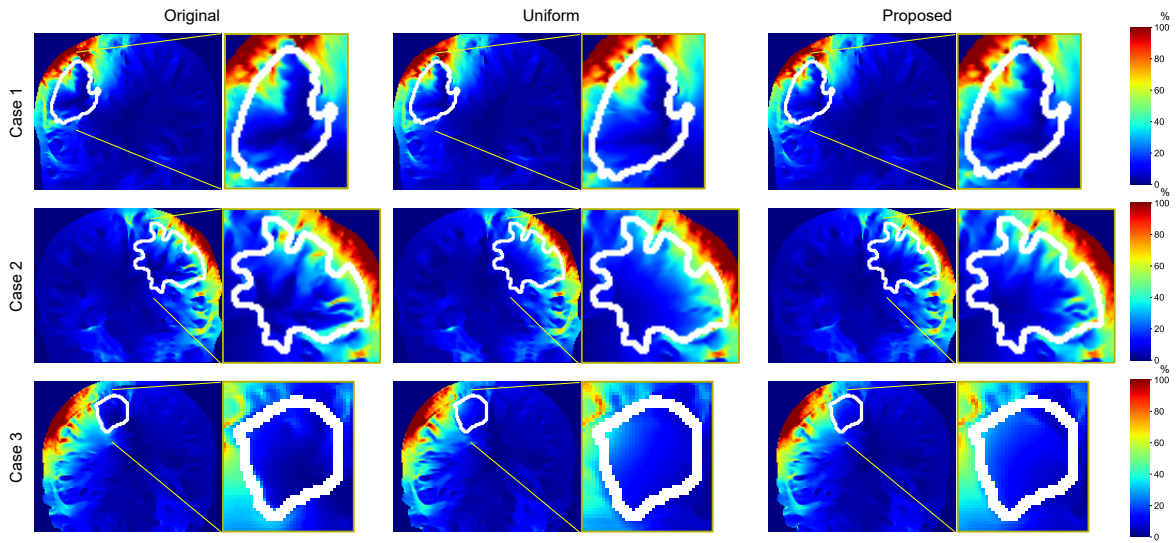

**FIGURE 5.** Electric field distribution for the three tumor cases under each method during TMS. White masks denote tumor regions.

**TABLE 6.** Summary of estimated conductivity values within the tumor region. Mean, SD, and range were calculated based on the case-wise mean conductivity values for each tumor subregion.

| Region | Method | Volume | | | Conductivity | | | Gradient | | |
|---|---|---|---|---|---|---|---|---|---|---|
| | | Mean | SD | Range | Mean | SD | Range | Mean | SD | Range |
| Rim | Original | | | | 0.335 | 0.105 | 0.169-0.569 | 0.144 | 0.061 | 0.042-0.278 |
| | Uniform | 9520 | 7739 | 995-28901 | 0.377 | 0.113 | 0.024-0.181 | 0.104 | 0.037 | 0.024-0.181 |
| | Proposed | | | | 0.366 | 0.106 | 0.132-0.512 | 0.100 | 0.034 | 0.014-0.152 |
| Core | Original | | | | 0.513 | 0.280 | 0.243-1.184 | 0.165 | 0.071 | 0.087-0.310 |
| | Uniform | 20546 | 26738 | 242-91258 | 0.404 | 0.098 | 0.263-0.594 | 0.081 | 0.042 | 0.020-0.157 |
| | Proposed | | | | 0.409 | 0.106 | 0.132-0.608 | 0.076 | 0.037 | 0.015-0.131 |
| Whole | Original | | | | 0.378 | 0.134 | 0.183-0.660 | 0.155 | 0.062 | 0.051-0.291 |
| | Uniform | 30067 | 34346 | 1237-120159 | 0.388 | 0.103 | 0.197-0.558 | 0.088 | 0.038 | 0.025-0.147 |
| | Proposed | | | | 0.392 | 0.105 | 0.141-0.542 | 0.084 | 0.033 | 0.014-0.133 |

**TABLE 7.** Summary of estimated conductivity values within the tumor region at 10 kHz, 1 MHz, and 100 MHz. Mean, SD, and range were calculated based on the case-wise mean conductivity values for each tumor subregion.

| Region | 10 kHz | | | 1 MHz | | | 100 MHz | | |
|---|---|---|---|---|---|---|---|---|---|
| | Mean | SD | Range | Mean | SD | Range | Mean | SD | Range |
| Rim | 0.366 | 0.106 | 0.132-0.512 | 0.525 | 0.188 | 0.276-0.866 | 0.739 | 0.128 | 0.619-1.063 |
| Core | 0.409 | 0.106 | 0.132-0.608 | 0.664 | 0.199 | 0.294-1.007 | 0.850 | 0.126 | 0.638-1.124 |
| Whole | 0.392 | 0.105 | 0.141-0.542 | 0.610 | 0.197 | 0.301-0.970 | 0.805 | 0.129 | 0.631-1.071 |

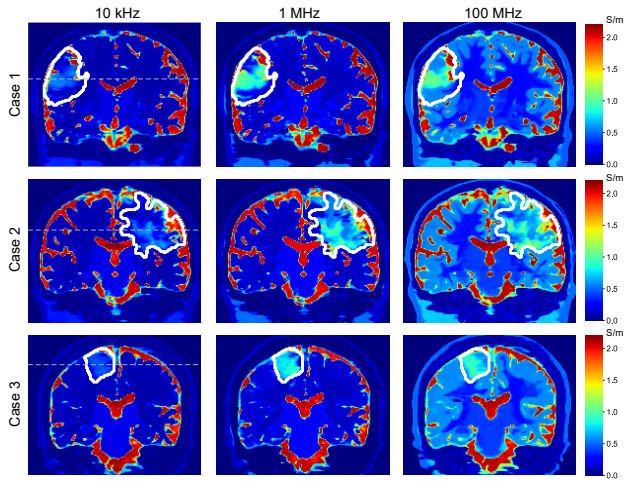

**FIGURE 6.** Estimated electrical conductivity distribution in brain tumors at 10 kHz, 1 MHz, and 100 MHz. White masks denote tumor regions.

The spatial variations in conductivity, as shown in Fig. 3 appear to correspond with tissue characteristics such as edema, necrosis, and rim-associated intensity gradients observed in T1- and T2-weighted MR images. Similar associations have been reported in previous studies [44], [45]. Lesbats et al. and Park et al. observed higher conductivity in the tumor rim than in the core in implanted gliomas in rats and mice [46], [47]. In contrast, Tha et al. found that in human gliomas, the tumor core exhibited higher conductivity than the rim [22], which aligns with our findings shown in TABLE 6 (rim < core). This discrepancy may stem from differences in the pathological structure and tissue response between implanted and naturally occurring tumors, which can influence conductivity distributions. Future studies should include a larger number of cases to facilitate a more comprehensive analysis of these trends. Furthermore, as shown in TABLE 4, a steeper conductivity gradient was observed in the rim region than in the core, likely reflecting the sharp transition in tissue properties at the tumor boundary. These findings suggest that, although the average conductivity tends to be lower in the rim than in the core, distinctive structural changes occur around the rim, which may aid in detecting abnormal tissue in the peritumoral region.

As demonstrated in Fig. 5, the conductivity maps generated using the proposed method are applicable to electric field simulations, further supporting the utility of our approach. Although the overall electric field distributions did not vary significantly among the three methods owing to the similar conductivity estimates for major tissues such as the CSF and skull from CondNet-TART, differences were evident within the tumor regions.

For example, in the original method, tumor voxels resembling CSF often received high conductivity values, leading to locally overestimated electric fields. Additionally, the conductivity in the tumor core was estimated to be unnaturally low, potentially impairing the accurate modeling of current propagation. In contrast, the proposed method yielded a smoother conductivity profile and suppressed the local field intensities within the tumors. These observations indicate that inaccurate conductivity estimation, particularly over- or underestimation within tumor tissues, can substantially alter local electric field distributions, potentially affecting the precision of stimulation targeting in clinical applications. This effect is likely to be especially critical when functional regions, such as the motor or language cortex, are located within or adjacent to the tumor, as it could impact the prediction of stimulation-induced responses [13], [14], [15]. Therefore, the proposed method, which enables a more accurate estimation of tumor conductivity, may contribute to more reliable preoperative simulation outcomes.

These results support the applicability of the proposed approach for patient-specific modeling of neuromodulation. Notably, the conductivity maps derived from standard T1- and T2-weighted MR images successfully captured tumor-specific heterogeneity, including edema, necrosis, and rim-associated gradients. Such heterogeneity is essential for realistic electric field simulations and has often been overlooked in previous models that assigned homogeneous values to the tumor regions. Notably, the significance of these contributions lies in their independence from specialized MRI sequences and manual annotation.

As shown in Fig. 6 and TABLE 7, the proposed method reproduced physiologically plausible tissue characteristics at 1 MHz and 100 MHz, comparable to those observed at 10 kHz, while also estimating conductivity values that reflected frequency-specific properties. This demonstrates that the method can adapt to different frequencies, extending its applicability to various neurostimulation modalities (e.g., TMS and TIS) and suggesting its potential for applications in the higher-frequency (MHz and above) range.

The ability of the proposed method to noninvasively estimate the conductivity distributions in brain tumors marks a significant advancement. This enables the image-based assessment of tumor electrical properties, which has been challenging to address. This capability enhances the accuracy of electromagnetic simulations and provides a valuable tool for enhancing our understanding of tumor pathophysiology, which is often complex and heterogeneous in nature. For example, the estimated conductivity maps could improve the precision of therapeutic approaches that depend on electrical properties, including radiation therapy [48] and radiofrequency ablation [49], [50], [51]. Additionally, extending dielectric property estimation to higher frequencies, although no experimental data are available, contributes to microwave-based applications such as microwave hyperthermia [52], [53], microwave ablation [54], [55], [56], microwave imaging [57], [58], assessment of human protection from microwaves [59], and other emerging microwave-based medical applications [60] by providing accurate tissue dielectric properties. Future work should aim to extend this method to a broader range of tumor cases and integrate it with diverse therapeutic modalities, thereby supporting more accurate and individualized strategies in clinical settings.

This study had several limitations. First, although only 12 tumor cases were available, the framework was pretrained on a large cohort of healthy participants (over 55), which enabled robust learning of head conductivity distributions. Therefore, the tumor dataset primarily served as a proof-of-concept to demonstrate its applicability to pathological conditions rather than to establish population-level statistics. Additional validation using a more diverse set of cases is essential to assess the generalizability and clinical applicability of the proposed method. Second, the estimated conductivity values were not directly compared with the ground truth data. Our current evaluation relies on visual assessments and statistical summaries, such as mean values and gradients. As the true conductivity distribution remains unknown, future studies should seek validation through direct measurements or independent verification using alternative methods to provide stronger evidence of their accuracy. Third, our validation was confined to frequencies up to approximately 100 MHz owing to the limited availability of measured dielectric property data; however, in principle, the framework can be extended to the GHz band once sufficient reference data become available [61], such as recent experimental measurements of human tissue dielectric properties [62].

## V. CONCULUSION
In this study, we established the first framework for the noninvasive and dispersive estimation of brain tumor conductivity. Our approach combines deep learning with a biophysically informed correction model using T1- and T2-weighted MRI to generate conductivity maps that capture intratumoral heterogeneity and inter-individual variability. The resulting maps demonstrated smooth, physiologically plausible spatial distributions and were directly applicable to the electric field stimulation analysis. Furthermore, the framework demonstrated voxel-wise dispersive mapping for estimating the conductivity at different frequencies. These advancements represent a significant step toward improving the precision of clinical neurostimulation and other microwave-based therapeutic applications in patients with brain tumors, such as hyperthermia and ablation.

## ACKNOWLEDGMENT
We thank ENAGO for English language editing support, and ChatGPT-5.0 (OpenAI) for assistance in language refinement.